\documentclass[17pt]{extarticle}
\usepackage{tikz}

\usepackage{amsfonts,amsmath}
\usepackage{extsizes}
\mathchardef\mhyphen="2D 
\usepackage[a4paper]{geometry}
\usepackage{cleveref}
\usepackage{microtype}
\usepackage[shortlabels]{enumitem} 
\usepackage{relsize}

\def\zo/{$0\mkern2mu\mhyphen1$}

\def\nn/{$n \times n$}

\title{On the Pernici-Wanless Expansion for the Entropy (and Virial Coefficients) of a Dimer Gas on an Infinite Regular Lattice}
\author{Paul Federbush \\ Department of Mathematics \\ University of Michigan \\
Ann Arbor, MI 48109-1043 \\
pfed@umich.edu}
\date{\today}

\usepackage{amsthm} 

\newtheorem{conj*}{Conjecture} 

\numberwithin{lemma}{section}
\numberwithin{conj*}{section}

\numberwithin{equation}{section}

\usepackage{listings}

\usepackage{xcolor}
\definecolor{codegreen}{rgb}{0,0.6,0}
\definecolor{codegray}{rgb}{0.5,0.5,0.5}
\definecolor{codeorange}{rgb}{1,0.49,0}
\definecolor{backcolour}{rgb}{0.95,0.95,0.96}

\lstdefinestyle{mystyle}{
    backgroundcolor=\color{backcolour},   
    commentstyle=\color{codegray},
    keywordstyle=\color{codeorange},
    numberstyle=\tiny\color{codegray},
    stringstyle=\color{codegreen},
    basicstyle=\ttfamily\footnotesize,
    breakatwhitespace=false,         
    breaklines=true,                 
    captionpos=b,                    
    keepspaces=true,                 
    numbers=left,                    
    numbersep=5pt,                  
    showspaces=false,                
    showstringspaces=false,
    showtabs=false,                  
    tabsize=2,
    xleftmargin=10pt,
    basicstyle = \large\ttfamily
}

\lstdefinelanguage{Maple}%
{morekeywords={and,assuming,break,by,catch,description,do,done,%
elif,else,end,error,export,fi,finally,for,from,global,if,%
implies,in,intersect,local,minus,mod,module,next,not,od,%
option,options,or,proc,quit,read,return,save,stop,subset,then,%
to,try,union,use,uses,while,xor,series,degree,convert,simplify,coeff,collect,normal,with,nops,partition,},%
sensitive=true,%
morecomment=[l]\#,%
morestring=[b]",%
morestring=[d]"%
}[keywords,comments,strings]%

\begin{document}

\maketitle
\begin{abstract}
	We work with the following expression for the entropy (density) of a dimer gas on an infinite $r$-regular lattice
	\begin{align*}
		\lambda(p) = \frac{1}{2} \left[ p\ln(r)-p\ln(p)-2(1-p)\ln(1-p)-p \right]+\sum_{k=2}^\infty d_kp^k
	\end{align*}
	where the indicated sum converges for density, p, small enough. 
	Pernici has computed the coefficients $d_k$ for $k \leq 12$. He
	found these $d_k$ to be polynomials in certain interesting ``geometric quantites" arising in the work of Wanless. Each of these quantities is the number density of isomorphic mappings of some graph into the lattice (graph). So for a bipartite lattice
	\begin{align*}
		d_2 &= c_2 \\
		d_3 &= c_3\\
		d_4 &= c_4 + c_5 \hat{G}_1 \\
		d_5 &= c_6 + c_7 \hat{G}_1.
	\end{align*}
	The $c_i$ depend only on $r$. Here $\hat{G}_1$ is the density of mapping classes of the four loop graph into the lattice. The limit of $1/n$ times the number of such mapping classes into a lattice of volume $V$ as $V$ goes to infinity. The infinite volume limit. $V$ equals $2n$
	equals the number of vertices.

	There is a simple linear relation that yields the $k^\text{th}$ virial coefficient from the value of $d_k$! 
	We feel this expression gives the deepest insight into the virial coefficients so far obtained.

	What we show in this paper is that such polynomial relations for the $d_k$ in these geometric quantities holds for the $d_k$ for $k \leq 27$. Of course we expect it to hold for all $k$. 
	We use the same computation procedure as Pernici. We note this procedure is not rigorously established. So far a procedure for the physicist, perhaps not the mathematician (their loss).
	It is a worthy challenge for the mathematical physicist to supply the needed rigor. 
\end{abstract}
\newpage

\section{Introduction}\label{sec:1}
We start with the following expansion for the entropy (density) of a dimer gas on an $r$-regular infinite lattice

	\begin{align} \label{1.1}
		\lambda(p) = \frac{1}{2} \left[ p\ln(r)-p\ln(p)-2(1-p)\ln(1-p)-p \right]+\sum_{k=2}^\infty d_kp^k
	\end{align}
	a rigorously established expression for hyper-rectangular lattices holding if the density, $p$, (twice the dimer density per vertex) is small enough. 
	This expansion was first written down by me, \cite{1}, from a clever, tricky, but non-rigorous development. 
	Surprisingly enough, there was an almost trivial, rigorous derivation, from a known rigorous formula, \cite{2},\cite{3}. 
	
	Pernici developed a procedure we call ``the high-$j$ limit" to extract expressions for the $d_k$, \cite{4}, from the work of Wanless, \cite{5}. 
	Both the works of Pernici and that of Wanless are highly non-trivial, the efforts of a master computational mathematical physicist and that of a master computational combinatorist balancing each other.

	We limit our discussion here to bipartite lattices. Using the high-$j$ limit procedure Pernici computed the $d_k$ for $k\leq 12$. They turned out to be polynomials in terms of geometric quantities, $\hat{G}_i$, from the work of Wanless. (Our notation differs from the notation in previous references.) Each $\hat{G}_i$ is the density of some isomorphic mapping class into the lattice. 
	
	So, $\hat{G}_1$ is the limit as $n$ goes to infinity of $\frac{1}{n}$ times the number of isomorphism classes of mappings of the $4$-loop graph into a $2n$ vertex approximation to the infinite lattice graph.
	Here $n$ is half the number of vertices.
	We take the volume of a region to be the number of vertices it contains. 
	The infinite volume limit is a place where mathematical rigor remains to be pursued. 
	The following four figures are the graphs being mapped into the lattice to define the first four $\hat{G}_i$. We have
	a similar limit as with $\hat{G}_1$ for $\hat{G}_i$ with $n$ going to infinity.

	
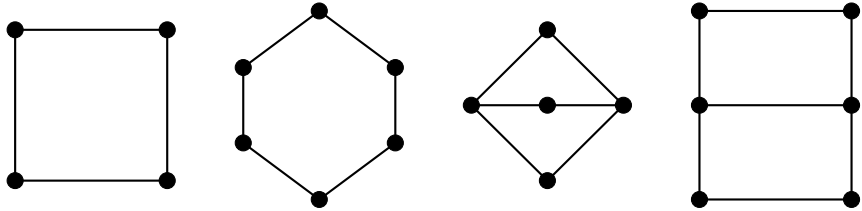
\begin{center}
		\centering
		\begin{tikzpicture}

			\draw[fill=black] (0,0) circle (3pt);
			\draw[fill=black] (0,2) circle (3pt);
			\draw[fill=black] (2,0) circle (3pt);
			\draw[fill=black] (2,2) circle (3pt);

			\draw[thick] (0,0) -- (2,0) -- (2,2) -- (0,2) -- (0,0);

			\draw[fill=black] (3,1.5) circle (3pt);

			\draw[fill=black] (3,0.5) circle (3pt);

			\draw[fill=black] (5,1.5) circle (3pt);

			\draw[fill=black] (5,0.5) circle (3pt);

			\draw[fill=black] (4,2.25) circle (3pt);

			\draw[fill=black] (4,-.25) circle (3pt);

			\draw[thick] (4,-.25) -- (3,0.5) -- (3,1.5) -- (4,2.25) -- (5,1.5) -- (5,0.5) -- (4,-0.25);


			\draw[fill=black] (6,1) circle (3pt);
			\draw[fill=black] (7,1) circle (3pt);
			\draw[fill=black] (7,2) circle (3pt);
			\draw[fill=black] (7,0) circle (3pt);
			\draw[fill=black] (8,1) circle (3pt);

			\draw[thick] (6,1) -- (7,2) -- (8,1) -- (7,0) -- (6,1);
			\draw[thick] (6,1) -- (7,1) -- (8,1);

			\draw[fill=black] (9,-0.25) circle (3pt);

			\draw[fill=black] (9,1) circle (3pt);

			\draw[fill=black] (9,2.25) circle (3pt);
			\draw[fill=black] (11,-0.25) circle (3pt);

			\draw[fill=black] (11,1) circle (3pt);

			\draw[fill=black] (11,2.25) circle (3pt);
			\draw[thick] (11,2.25)--(11,1)--(11,-0.25)--(9,-0.25)--(9,1)--(9,2.25);
			\draw[thick] (11,2.25) -- (9,2.25);
			\draw[thick] (9,1) -- (11,1);
		\end{tikzpicture}
	\end{center}

We take the expressions for the $d_k$ from eq.(41) of \cite{4} where there are notational changes made, and more important to notice, eq.(40) of \cite{4} differs from our \cref{1.1}. 

	\begin{align}
		d_2 &= c_1 \\
		d_3 &= c_2\\
		d_4 &= c_3 + c_4 \hat{G}_1 \\
		d_5 &= c_5 + c_6 \hat{G}_1 \\
		d_6 &= c_7 + c_8 \hat{G}_1 + c_9 \hat{G}_2 + c_{10} \hat{G}_3 \label{1.6}\\
		d_7 &= c_{11} + c_{12} \hat{G}_1^2 + c_{13} \hat{G}_1 + c_{14} \hat{G}_2 +c_{15} \hat{G}_3 + c_{16}\hat{G}_4.
	\end{align}

	From the fact that $c_1$ through $c_6$ are positive, one gets as pointed out in \cite{4} the positivity of $d_2$ through $d_5$. Note the $c$'s only depend on $r$. 
	These positivities are all emphasized in \cite{6}. It is disappointing that the beautiful geometric representation with the $\hat{G}_i$ does not immediately yield more positivities of the $d_k$. 
	One can express the coefficients $m_k$ of the virial expansion 
	\begin{align}
		P(p) = p/2 + \sum_{k=2}^\infty m_k p^k
	\end{align}
	as follows, from eq.(12) of \cite{7}
	\begin{align}
		m_k = (k-1)\left( \frac{1}{(k)(k-1)}-d_k \right).
	\end{align}
	One should note that eq.(8) and eq.(9) of \cite{7} agrees with our \cref{1.1} with their $a_k$ replaced by $d_k$. 
	
	A small note, our extension of the result of Pernici, from 12 to 27 for bipartite lattices, is instead from 12 to 23 for general $r$-regular lattices and from 12 to 32 for the hexagonal lattice. 

	Section 2 discusses the high-$j$ limit, which is intimately connected to the infinite volume limit. Section 3 treats the theory of our extension of Pernici's result. Indicating the line of extension of the method of \cite{8} to the current problem. One finds the current paper basically almost a corollary of \cite{8}.

	We leave to the mathematical physicists younger and smarter than we the following important and difficult problems. There is much work to be done here.

\begin{enumerate}
	\item Prove the validity of the ``high-$j$ limit", for hyper-rectangular lattices, and for general $r$-regular lattices. It seems to me likely that one must
	restrict oneself to vertex-transitive lattices, see \cite{9}, and the work of Csikvari therein.

	\item Extend the results of \cite{8} from $k = 27$ to all values of $k$. Here we will want to extend the generalization of the problem in \cite{8} to the problem herein. That is the result in Appendix B.1. 
	\item Use the expressions of the $d_k$ and $m_k$ in terms of the $\hat{G}_i$ to obtain further positivity results.
\end{enumerate}

\section{The High-$j$ Limit}\label{sec:2}

We deal with an infinite $r$-regular lattice. We have a sequence of finite graphs, finite approximations to the infinite lattice, the $i^\text{th}$ approximate graph with $2n_i$ vertices. Recall the volume of the $i^{\text{th}}$ graph is by definition $2n_i$. For a hyper-rectangular lattice, an ideal finite graph approximation would be a hyper-cube of volume $2n_i$, and with periodic boundary conditions. The density, $p_i$ is chosen to satisfy 
\begin{align} \label{2.1}
	j_i = p_in_i
\end{align}
where we're dealing with $j_i$-matchings. That is, there are $j_i$ dimers on the finite graph approximation. $p_i$ is the number of vertices covered by the dimers per unit volume. $p$ is an intensive quantity and ideally it would be fixed at the infinite volume limit, but $j$ and $n$ must be integers, so the value of $p$ must vary from the infinite volume limit. (This is a standard nuisance in the study of the infinite volume limit.)

Our treatment of the high-$j$ limit is based on Pernici's work in \cite{4} and the developments in \cite{8}. It may be said to be our interpretation of the limit taken by Pernici. 
We find it easier to follow than Pernici's treatment, but it makes no direct contribution to understanding why the limit gives correct answers.

Following eq.(8) and eq.(10) of \cite{4} we let $m(j_i)_{n_i}$ be the number of $j_i$ matchings on the $i^{\text{th}}$ approximating graph of volume $2n_i$. 
From eq.(11) and eq.(12) of \cite{4} and eq.(3.11) and eq.(3.12) of \cite{8} we get the expression
\begin{align} \label{2.2}
	m(j_i)_{n_i} = \left( \dfrac{n_i^{j_i}}{j_i!}\right)\ln \left( \sum_{h \geq 0} \dfrac{\hat{a}_h(j_i,\{\epsilon_t\})}{n_i^h}\right).
\end{align}

We have suppressed dependencies on $r$. There is the important subtlety that the $\epsilon_t$ may depend on $n_i$!

We note the basic thermodynamic limit as 
\begin{align} \label{2.3}
	m(j_i)_{n_i} \sim e^{2n_i \lambda(p)}
\end{align}
as $n_i$ goes to infinity. More precisely
\begin{align} \label{2.4}
	\lim_{n_i \to \infty} \frac{1}{2n_i} m(j_i)_{n_i} = \lambda(p).
\end{align}

If we denote this equation in paragon form
\begin{align} \label{2.5}
	\lim_{\substack{n \to \infty \\ p \text{ fixed}}} L(j,n) = R(p) 
\end{align}
where $L$ and $R$ are the left and right side functions in \cref{2.4} and we remember \cref{2.1} as 
\begin{align} \label{2.6}
	j = pn.
\end{align}

Then the high-$j$ limit replaces \cref{2.5} by 
\begin{align} \label{2.7}
	\lim_{j \to \infty} \dfrac{ \left[n^{-h}\right] (2nL(j,n))}{\left[n^{-h}\right](2nR(j,n))} = 1.
\end{align}

As in \cite{4} and \cite{8}, $[n^{-h}]f(n) = a_n$ if $f(n) = \sum_{k} \frac{a_k}{n^k}$.

The numerator and the denominator of the quantity inside the limit in \cref{2.7} are both polynomials in $j$. Their degrees must be equal therefore. 
It is easy to see the denominator polynomial degree is $h+1$. Therefore,
\begin{align} \label{2.8}
	\left[j^k \frac{1}{n^h} \right]\left(\frac{n^j}{j!} \right) \ln\left( \sum_{h \geq 0} \dfrac{\hat{a}_h(j,\left\{\epsilon_t\right\})}{n^h}\right)=0 \quad k \geq h+2.
\end{align}
where $\left[j^k \frac{1}{n^h}\right]$ is a generalization of $\left[ \frac{1}{n^h} \right]$. The treatment of the $k = h+1$ term is left to the next section.

\section{Computational Matters} 

\subsection{On $\hat{a}_n$} \label{sec:3}

The technical improvement we introduce into the treatment of Pernici in \cite{4} is the set of quantities $\hat{a}_n$, from \cite{8}. eq.(2.8) above is a generalization of eq.(16) of \cite{4}, which is the statement for an idealized problem where one is dealing with the Bethe lattice (a problem idealized out of reality). That is, $a_n$ of \cite{4} are the $\hat{a}_n$ for the Bethe lattice.

We have to make modifications of the line of proof in parts of \cite{8}. In particular, by eq.(3.10) of \cite{8} the $\epsilon_i$ therein are intensive (not extensive) variables. 
For us the $\hat{G}_i$ are intensive but the $\epsilon_i$ are \emph{extensive} variables! 

What we need from the reader so that he or she understand this section, is that the proof in Appendix B of \cite{8} is understood. 
Results and definitions from other places may be referred to, but the proofs in other places need not be understood. 
But the proof in Appendix B is tricky, subtle, and in short requires a serious effort to master. (A personal aside. After more than a year I went back to look at this section and misunderstood its cleverness and subtlety, and ended up putting a number of truly ignorant ``corrections" to this section on the web. Make sure you have the latest version! In my defense, I am blessed, or cursed, with a very poor memory -- one I have always had, not just with age. I am 87.)

In novel form, I center this section around a Maple program (using the 2018 or 2019 version) used in the computer proof of Lemma B.1 of \cite{8}. 
I believe it would not be difficult for the reader familiar with say Mathematica instead to transcribe. One goes through the same steps in
unravelling the action of the Maple program as in developing a prose proof.

\newpage

\subsection{Main Algorithm Code}

\begin{lstlisting}[language=Maple, caption={The Program.},label={lst:code}, mathescape=true, breaklines=true]
mm:=10;
lL:=3;
aQ:=0;
rh:=1;
jq:=2;

for vq from 1 to mm do:
u||vq:=vq*(1-((1/(10+aQ))*vq)):
od:

sc||1:=j:
for qq from lL to mm do:
sc||qq:= (j-qq+1)*(sc||(qq-1)):
od:

for ih from 2 to lL-1 do:
ch||ih:=0:
od:

with(combinat,partition):
pp:=partition(mm):
ww:=nops(pp):

for qQ from 1 to ww do:
ww|qQ:=nops(qQ,ww):
od:

for rQ from 1 to ww||qQ do:
nn||(op(rQ,op(qQ,ww))):=nn||
(op(rQ,op(qQ,ww))) +1:
od:

for iq from lL to ww do:
for kq from lL-1 to iq -1 do:

read ``/Users/pfed/Box sync/LSA-Files/88-Good/'';

aa:=1:
bb:=1:
for s from 1 to mm do:
aa:=aa*((u||(s+1))^(nn||s))/((nn||s)!):
bb:=bb*((rh*ch||s)^(nn||s))/((nn||s)!):
od:

wwrhjq||
iq:=degree(coeff((collect(FF||iq,j,normal),rh^jq)),j):

od:

F||iq:=aa:
ee||iq:=bb:

F||iq:=simplify(F||iq):
ee||iq:=simplify(ee||iq):

FF||iq:=F||iq:
for qQ from lL to mm do:
FF||iq:=FF||iq +(ee||qQ)*(sc||qQ)*(z^qQ)*(1+subs(j=j-qQ,F||iq)):
od:

FF||iq:=series(FF||iq,z=0,mm+1):
FF|iq:=convert(FF|iq,polynom):
od:

wwrhjq||
iq:=degree(coeff((collect(FF||iq,j,normal),rh^jq)),j);


\end{lstlisting}

\subsection{Subroutine Code}

\begin{lstlisting}[language=Maple, caption={The Subroutine.},label={lst:subcode}, mathescape=true, breaklines=true]
Sq2:={}:

if wwrhjq||iq > $100^2\times (lL-1)$ then:
if wwrhjq||KQ < $(100)^2\times lL$ then:

Sq2:= Sq2 union{wwrhjq||kq}
	union{wwrhjq||KQ}:
	
fi:
fi:
od:


\end{lstlisting}

\end{document}